**Nonvolatile SRAM architecture using MOSFET-based spin-transistors**

Yusuke Shuto[1,4 a)], Shuu'ichirou Yamamoto[2,4], and Satoshi Sugahara[1,3,4 b)]

[1]Imaging Science and Engineering Laboratory, Tokyo Institute of Technology, 4259-G2-14 Nagatsuta, Midori-ku, Yokohama 226-8502, Japan
[2]Department of Information Processing, Tokyo Institute of Technology, 4259-G2-28 Nagatsuta, Midori-ku, Yokohama 226-8502, Japan
[3]Department of Electronics and Applied Physics, Tokyo Institute of Technology, 4259-G2-14 Nagatsuta, Midori-ku, Yokohama 226-8502, Japan
[4]CREST, Japan Science and Technology Agency, 4-1-8 Honcho, Kawaguchi-shi, Saitama 332-0012, Japan

The authors proposed and computationally analyzed nonvolatile static random access memory (NV-SRAM) architecture using metal-oxide-semiconductor field-effect transistor (MOSFET) type of spin-transistors referred to as pseudo-spin-MOSFET (PS-MOSFET). PS-MOSFET is a new circuit approach to reproduce the functions of spin-transistors, based on recently progressed magnetoresistive random access memory (MRAM) technology. The proposed NV-SRAM cell can be simply configured by connecting two PS-MOSFETs to the storage nodes of a standard SRAM cell. The logic information of the storage nodes can be electrically stored into the magnetic tunnel junctions (MTJs) of the PS-MOSFETs by current-induced magnetization switching (CIMS), and the stored information is automatically restored when the inverter loop circuit wakes up. In addition, the proposed NV-SRAM cell has no influence on the performance of normal SRAM operations. Low power dissipation and high degree of freedom of MTJ design are also remarkable features for NV-SRAM using PS-MOSFETs.

a) Electronic mail: shuto@isl.titech.ac.jp

b) Electronic mail: sugahara@isl.titech.ac.jp

Over recent years, it has been well recognized for next-generation advanced complementary metal-oxide-semiconductor (CMOS) logic circuits that static power dissipation during their standby mode becomes a serious inevitable problem. Static power dissipation in CMOS circuits is caused by large leakage currents of their constituent CMOS transistors, which is related to physical phenomena in highly-scaled and largely-integrated advanced CMOS devices. Power gating is expected to be the most promising architecture to reduce static power, in which logic circuits are divided into the so-called power domains and each power domain can individually execute power management (frequent shutdown) using a sleep transistor connected to it.[1-3] Static random access memory (SRAM) and flip-flop (FF) are used for data latches in high speed logic such as a microprocessor. However, these memory devices cannot be shut down without losing their information. In order to establish ideal power gating system, important logic information in the power domains should be stored on the spot without data transfer in view of its circuit performance. Therefore, nonvolatile SRAM (NV-SRAM) and nonvolatile FF (NV-FF)[4-9] are required for idealized power gating system.[10-12]

NV-SRAM can be configured by connecting magnetic tunnel junctions (MTJs)[13-15] to the storage nodes of a standard SRAM cell.[7,8] Although these NV-SRAM cells have to use a magnetic field for magnetization switching of MTJs, current-induced magnetization switching (CIMS) architecture is attractive for fully electrical control of magnetization configurations (resistance states) of MTJs[16,17] and thus is suitable for NV-SRAM. By introducing CIMS technology with a specific circuit configuration for the connection of MTJs, fully electrical store/restore operations are possible for NV-SRAM.[10,11] Although occupied area of such a NV-SRAM cell using MTJs with CIMS architecture could be designed to be close to that of a standard SRAM cell, the MTJs connected to the cell would deteriorate the performance of normal SRAM operations owing to unwanted currents passing through the MTJs.[10] Therefore, there exists tight restriction for the degree of freedom of MTJ design to achieve normal SRAM operations with high performance.[12] In order to overcome the issues in NV-SRAM using MTJs with CIMS architecture, we proposed a spin-transistor approach.[10] In this paper, a NV-SRAM cell using pseudo spin metal-oxide-semiconductor field-effect transistors (PS-MOSFETs)[18] is proposed and computationally analyzed.

The proposed NV-SRAM cell can be configured by replacing MTJs in a previously demonstrated NV-SRAM cell[10,11] with MOSFET-based spin-transistors, such as spin-MOSFETs[19,20] or PS-MOSFETs,[18] as an active nonvolatile storage element. In practice, the usage of PS-MOSFETs is promising, since they can be easily configured by an ordinary MOSFET and MTJ using recently progressed magnetoresistive random access memory (MRAM) technology. Figure 1(a) shows the circuit configuration of a PS-MOSFET. A MTJ connected to the source of a MOSFET feeds back its voltage drop to the gate. The degree of negative feedback depends on the resistance states of the MTJ, and thus the actual input (gate-source) bias $V_{GS0}$ can be varied by the magnetization configurations of the MTJ even under a constant gate bias ($V_G$) condition. Therefore, the PS-MOSFET can possess high and low current drivabilities that are controlled by the magnetization configurations of the MTJ. In addition, CIMS for the MTJ can be controlled by $V_G$, as discussed later. Thus, the PS-MOSFET can reproduce the spin-transistor behavior of spin-MOSFETs and would be the most promising spin-transistor based on present MRAM technology.[16,17,21,22]

Figure 1(b) shows the circuit configuration of the proposed NV-SRAM cell using PS-MOSFETs. The NV-SRAM cell consists of a standard SRAM cell portion [two cross-coupled inverters ($INV_1$ and $INV_2$) with two data-access pass transistors] and additional two PS-MOSFETs (PM$_1$ and PM$_2$) connected to the two storage nodes (*Q* and *QB*). Each PS-MOSFET is composed of an ordinary MOSFET and MTJ, as described above. The store/restore (*SR*) line is shared by the gate terminals of PM$_1$ and PM$_2$, and it is used to turn on the PS-MOSFETs for store/restore operations. The control (*CTRL*) line is shared by the source terminals of PM$_1$ and PM$_2$, and it is used to control polarity (direction) of current passing through PM$_1$ and PM$_2$ for store operation. The connecting arrangement of the MTJs depends on power gating architecture (i.e., virtual ground or virtual $V_{DD}$). In the virtual ground (virtual $V_{DD}$) case, the free (pinned) layer of the MTJs in PM$_1$ and PM$_2$ is connected to the *CTRL* line and their pinned (free) layer is connected to the MOSFETs in PM$_1$ and PM$_2$. In this NV-SRAM, the two PS-MOSFETs are electrically separated from the standard SRAM portion by switching them to their off-state during normal SRAM operation mode.

Circuit operations of the NV-SRAM cell using PS-MOSFETs were analyzed by HSPICE program with our developed MTJ model including CIMS.[10] A 70nm CMOS process model[23] was used for the following simulations. The gate length (L) / width (W) of n-channel and p-channel MOSFETs were set to 70nm / 350nm, respectively. This relatively wide gate width was used readily to achieve CIMS. Although the gate width can be reduced by optimization of circuit design, significant reduction of the gate width requires reduction of critical current density $J_C$ for CIMS. This is the same as the situation of MRAM using CIMS architecture (the so-called spin transfer torque RAM; SPRAM or STT-RAM).[16,17] The supply voltage $V_{DD}$ for the cell was set to 1.5V. The MTJ model can well reproduce electrical properties of recently developed MgO-based MTJs, i.e., nonlinear tunneling-type (Simon's formula type) current-voltage characteristics in antiparallel magnetization and ohmic-like current-voltage characteristics in parallel magnetization.[15] A junction resistance $R_P(0)$ of the MTJ model in parallel magnetization at the zero bias voltage was varied from 5kΩ to 100kΩ, and a junction resistance $R_{AP}(0)$ of the MTJ model in antiparallel magnetization at the zero bias voltage is set to achieve a desired tunnel magnetoresistance ratio (TMR) (= $[R_{AP}(0)-R_P(0)]/R_P(0)$). A characteristic voltage $V_{half}$ (that is a voltage drop of MTJ when its TMR decreases to the half maximum) was varied from 0.3 to 1.0 V. A critical current $I_C$ for CIMS from parallel to antiparallel magnetization was designed by the relation of $I_C = V_C^{P \rightarrow AP}/R_P(0)$ using a critical bias voltage $V_C^{P \rightarrow AP}$. The same $I_C$ value was used for CIMS from antiparallel to parallel magnetization for simplicity. Thus, its critical bias voltage is given by $V_C^{AP \rightarrow P} = I_C \cdot R_{AP}(V_C^{AP \rightarrow P})$. In the following simulations, $V_C^{P \rightarrow AP}$ was set to a slightly small value of 0.5 V (Ref. 24) so that CIMS from antiparallel to parallel can occur within $V_{DD}$, i.e., $V_C^{AP \rightarrow P} < V_{DD}$. In our simulation, $I_C$ was varied from 100 μA to 5 μA, owing to changes in parameterized $R_P$. When $J_C$ is assumed to be $1 \times 10^6$ A/cm$^2$, its junction area is in a range between $1 \times 10^4$ nm$^2$ and $5 \times 10^2$ nm$^2$.

Figure 2(a) shows output characteristics of a PS-MOSFET with $R_P$ = 5 kΩ, TMR = 100%, and $V_{half}$ = 0.5 V, where $V_G$ increases from 0 to 1 V in steps of 0.2 V. It can be clearly seen that when the magnetization configuration of the MTJ is parallel (antiparallel), the current drivability is high (low), which is virtually the same as the spin-transistor behavior of spin-MOSFETs. Figure 2(b) shows CIMS behavior of the PS-MOSFET. By pulling up $V_G$ to 1.5 V, the PS-MOSFET can drive currents above $I_C$. CIMS from parallel to antiparallel magnetization can be achieved with a positive drain bias voltage, as shown in the first quadrant of the figure, and also CIMS from antiparallel to parallel magnetization can be achieved by applying reverse drain bias, as shown in the third quadrant of the figure.

PS-MOSFET is a new circuit approach to reproduce the functions of spin-MOSFETs, based on recently progressed MRAM technology.[16,17,21,22]

In the proposed NV-SRAM cell, the storage nodes $Q$ and $QB$ of the cross-coupled inverters have the two output voltages of high (H) and low (L) levels that correspond to the logic information of "1" and "0", respectively. In the store operation stage, the information of the storage nodes can be electrically stored in MTJ$_1$ and MTJ$_2$ as a resistance condition by CIMS, using a pulse signal of $V_{DD}$ applied on the *CTRL* line. The sequence of the store operation includes only 3 steps. Initially, both the *SR* and *CTRL* lines are at the L level. In the first step, the PS-MOSFETs are turned on by pull-up of the *SR* line to $V_{DD}$. In the second step, the *CTRL* line is also activated to $V_{DD}$. In the last step, the voltage levels of both the *SR* and *CTRL* lines are returned to the L level. Here, we consider the case that the node voltages $V_Q$ and $V_{QB}$ of the nodes $Q$ and $QB$ are at the H and L levels, respectively. During the first step, $V_Q$ = H is stored as the $R_{AP}$ state of MTJ$_1$. When the magnetization configuration of MTJ$_1$ is parallel, the resulting positive drain bias for PM$_1$ induces CIMS and it changes the magnetization configuration of MTJ$_1$ from parallel to antiparallel magnetization. In the case of antiparallel magnetization of MTJ$_1$, the magnetization configuration of MTJ$_1$ remains as it was before (see the first quadrant of Fig. 2(b)). During the second step, the information of $V_{QB}$ = L is stored as the $R_P$ state of MTJ$_2$, since the resulting reverse drain bias for PM$_2$ changes or holds the magnetization configuration of MTJ$_2$ as shown in the third quadrant of Fig. 2(b), and thus MTJ$_2$ possesses the parallel magnetization configuration regardless of its initial state. When $V_Q$ = L and $V_{QB}$ = H, the store operation can be done by the same manner. As a result, one of the MTJs connecting to the H level node is in the high resistance state ($R_{AP}$), and another MTJ connecting to the L level node is in the low resistance state ($R_P$), after the store operation.

The restore operation utilizes difference in the current drivability between $PM_1$ and $PM_2$.    After the store operation described above, $MTJ_1$ is set to the $R_{AP}$ state and $MTJ_2$ the $R_P$ state.    Before the supply voltage $V_{supply}$ of $INV_1$ and $INV_2$ begins to pull up, $PM_1$ and $PM_2$ are turned on by applying signal to the $SR$ line in advance.    By sweeping $V_{supply}$, the storage nodes $Q$ and $QB$ are electrically charged firstly owing to the parasitic capacitance effect of the cell and secondly owing to the p-channel MOSFETs of $INV_1$ and $INV_2$.    These nodes are simultaneously discharged by $PM_1$ and $PM_2$.    The node $Q$ is discharged more slowly than the node $QB$, since the current drivability of $PM_1$ is lower than that of $PM_2$ owing to the $R_{AP}$ and $R_P$ states of $MTJ_1$ and $MTJ_2$, respectively.    Therefore, $V_Q$ is higher than $V_{QB}$.    Although this difference between $V_Q$ and $V_{QB}$ ($V_Q > V_{QB}$) is tiny at the initial stage of the restore operation, it is enhanced by the push-pull operation between $INV_1$ and $INV_2$, after the n-channel MOSFET of $INV_1$ is turned on (This is caused by $V_Q >$ its threshold voltage $V_{thn} > V_{QB}$).    The resulting bistable condition represents the previously stored information in $MTJ_1$ and $MTJ_2$.    When $V_{supply}$ reaches to $V_{DD}$, the node information ($V_Q = H$ and $V_{QB} = L$) suspended at the previous normal SRAM operation mode can be completely restored.    After the information is recovered, $PM_1$ and $PM_2$ can be turned off in order to cut off unwanted leakage currents passing through the PS-MOSFETs during normal SRAM operations.

Figure 3(a) shows transient response in the node voltage $V_Q$ and $V_{QB}$ during the restore operation, where $MTJ_1$ and $MTJ_2$ are in the antiparallel and parallel magnetization configurations, respectively.    In our calculation, a sweep speed of the $SR$ line voltage $V_{SR}$ was set to 7.5V/nsec, and that of $V_{supply}$ 1.5V/nsec.    $V_{SR}$ pulls up to turn on $PM_1$ and $PM_2$ before $V_{supply}$ starts to be swept. $T_{active}$ represents a period while PM$_1$ and PM$_2$ are fully turned on as shown in the figure. When $T_{active}$ is enough long (e.g., $T_{active}$ = 2.8 nsec), the logic information stored in MTJ$_1$ and MTJ$_2$ are restored as a bistable condition of the nodes $Q$ and $Q_B$. However, $V_Q$ is dropped down from $V_{DD}$ owing to the effect of an undesired current through PM$_1$, as shown in the figure. This is the same as the problem of NV-SRAM using only MTJs,[10,11] as described previously. Nevertheless, the remarkable point is that $V_Q$ and $V_{QB}$ can establish an initial bistable condition, even when $V_{supply}$ increases slightly (see $V_Q$ and $V_{QB}$ after about 1.5 nsec from $V_{SR}$ turn-on). Since currents passing through PM$_1$ and PM$_2$ for the restore operation are not required after the initial bistable condition is achieved, PM$_1$ and PM$_2$ can be turned off before $V_{supply}$ reaches to $V_{DD}$, as shown by the case of $T_{active}$ = 1.4 nsec in the figure. Figure 3(b) shows currents passing through PM$_1$ during $T_{active}$. The undesired excess currents can be effectively cut off by shortening $T_{active}$.

Figure 4(a) shows transient response in the node voltages $V_Q$ and $V_{QB}$ with $T_{active}$ = 1.4 nsec, where $R_P$ is varied from 5 kΩ to 100 kΩ. The initial bistable state of the inverter loop can be established in the wide range of $R_P$. Figure 4(b) shows the power dissipation of MTJ$_1$ as a function of $T_{active}$ with various $R_P$ values during the restore operation. The power dissipation at MTJ$_1$ increases with decreasing $R_P$. However, it can be effectively reduced by shortening $T_{active}$. The lower limit of $T_{active}$ must be designed so that the initial bistable condition can be certainly established. Therefore, the power dissipation due to leakage currents through MTJ$_1$ and MTJ$_2$ can be suppressed by optimizing $T_{active}$ to a possibly low value, even when low resistive MTJs that are preferable to CIMS are used for PM$_1$ and PM$_2$. This feature is one of important advantages of the presented NV-SRAM cell using PS-MOSFETs in comparison with NV-SRAM using only MTJs in which highly resistive

MTJs must be used to suppress power dissipation during restore and normal SRAM operations.

The influence of $V_{half}$ and TMR variations on the restore operation was also analyzed. Their influence was confirmed to be minor, and thus MTJs with moderate $V_{half}$ (= 100 mA) and TMR (= 100%) are sufficient for the NV-SRAM operations. This is another feature for the bistable circuit configuration of our NV-SRAM. Requirements for $V_{half}$ and TMR are not so strict, which is quite different from the case of MRAM or SPRAM.

In summary, we proposed and computationally analyzed NV-SRAM architecture using PS-MOSFETs. PS-MOSFET is a new circuit approach to reproduce the functions of spin-transistors, based on recently progressed MRAM technology.[16,17,21,22] The proposed NV-SRAM cell can be simply configured by connecting two PS-MOSFETs to the storage nodes of a standard SRAM cell. The logic information of the storage nodes can be electrically stored into the MTJs of the PS-MOSFETs by CIMS, and the stored information is automatically restored when the inverter loop circuit wakes up. In addition, the proposed NV-SRAM cell has no influence on the performance of normal SRAM operations. Low power dissipation and high degree of freedom of MTJ design are also remarkable features for NV-SRAM using PS-MOSFETs.

**Figure captions**

FIG. 1.    (a) Circuit configuration of PS-MOSFET that consists of an ordinary n-MOSFET and MTJ.    (b) Circuit configuration of a proposed NV-SRAM cell using two PS-MOSFETs.

FIG. 2.    (a) Simulated output characteristics of PS-MOSFET with MTJ parameters of $R_P$ = 5 k$\Omega$, TMR = 100%, and $V_{half}$ = 0.5 V.    (b) CIMS behavior of PS-MOSFET.

FIG. 3.    (a) Transient behavior of $V_Q$ and $V_{QB}$ during the restore operation, where MTJ$_1$ and MTJ$_2$ are in the antiparallel and parallel magnetization configurations, respectively.    (b) Current $I_{MTJ1}$ passing through MTJ$_1$ in PS-MOSFET$_1$ during $T_{active}$.

FIG. 4.    (a) Influence of $R_P$ on the transient behavior of $V_Q$ and $V_{QB}$, where $R_P$ is varied from 5 k$\Omega$ to 100 k$\Omega$.    (b) Power dissipation of MTJ$_1$ as a function of $T_{active}$ with various $R_P$ values during the restore operation.

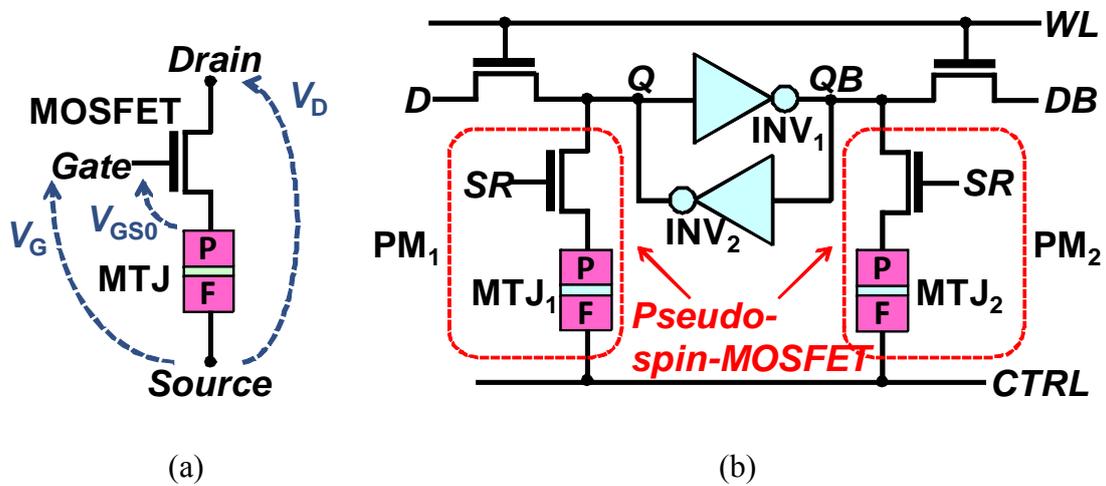

(a)                              (b)

Fig. 1.   <Color online>   Shuto, *et al.*

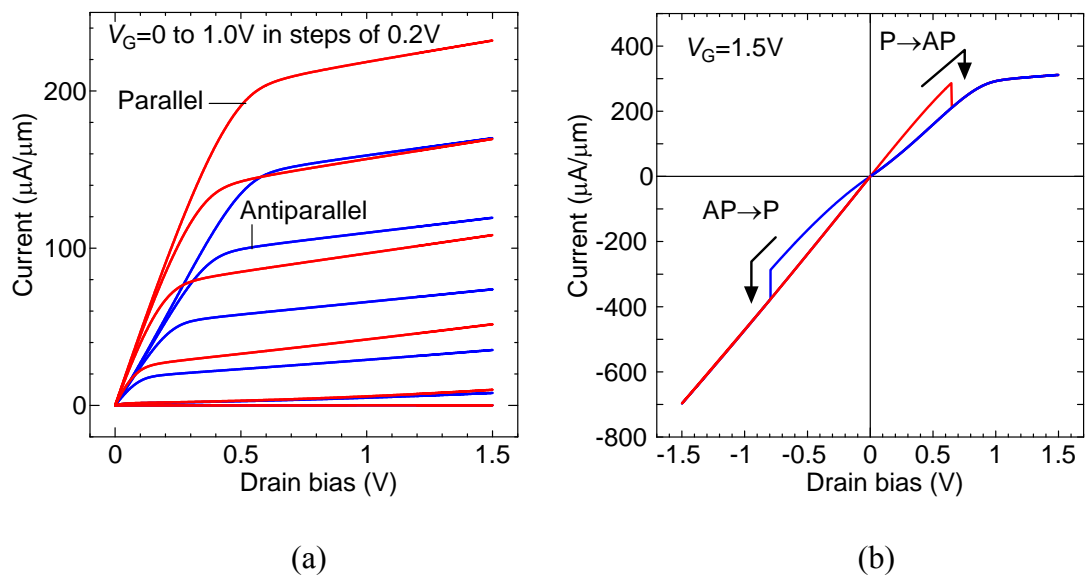

(a)

(b)

Fig. 2.    <Color online>    Shuto, *et al*.

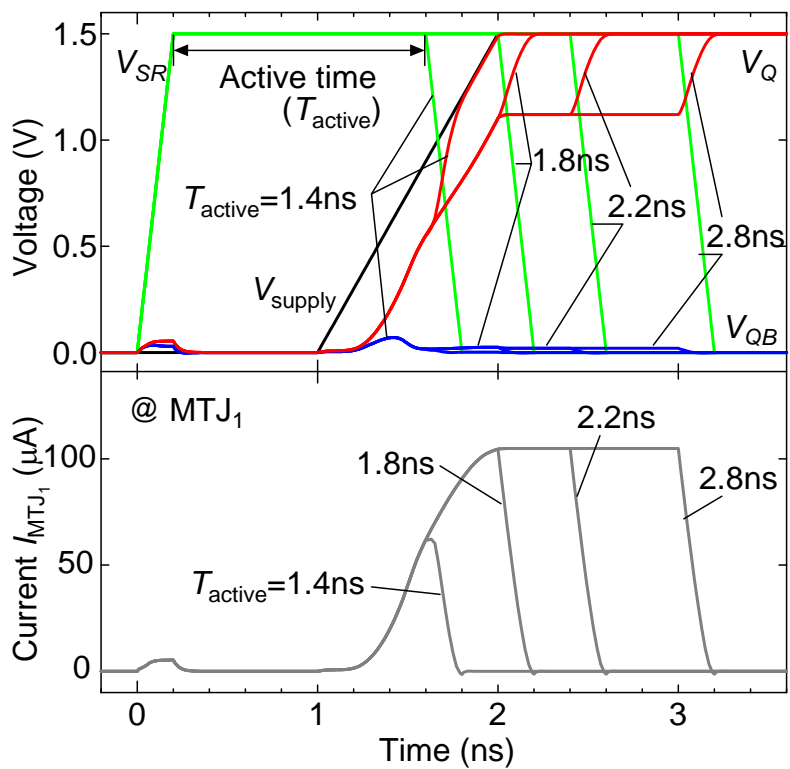

Fig. 3.   <Color online>   Shuto, *et al*.

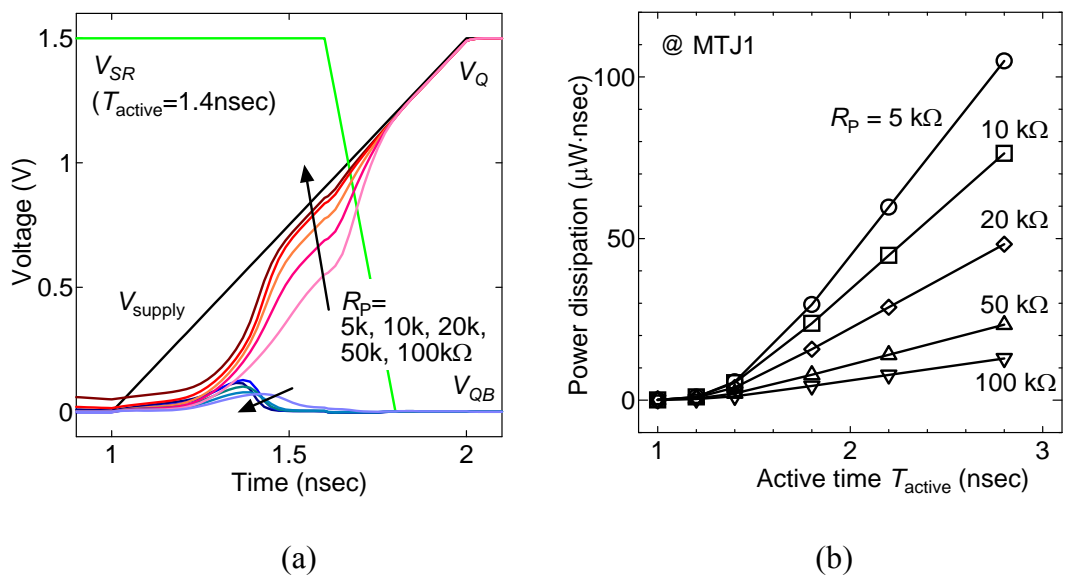

(a)

(b)

Fig. 4.    <Color online>    Shuto, *et al.*